\documentclass[12pt]{article}
\usepackage[colorlinks=true,linkcolor=blue,urlcolor=blue,filecolor=black,citecolor=red,pdfstartview=FitV,pdftitle={},pdfsubject={},
pdfkeywords={},pdfpagemode=None,bookmarksopen=true]{hyperref}
\usepackage{graphicx}
\usepackage{epstopdf}%
\usepackage{amsmath}
\usepackage{amsfonts}
\usepackage{amssymb}
\usepackage{color}%
\usepackage{dcolumn}
\usepackage{slashed}
\usepackage{amssymb,ulem}
\usepackage{float}
\usepackage{amsthm,amsmath,amssymb}
\usepackage{mathrsfs}
\usepackage{subfigure}
\usepackage{wrapfig}
\usepackage{indentfirst}
\providecommand{\U}[1]{\protect\rule{.1in}{.1in}}

\begin{document}

\vspace{12mm}

\begin{center}
{{{\Large {\bf Scalarization of Kerr-Newman black holes in the Einstein-Chern-Simons-scalar theory}}}}\\[10mm]

{Kun-Hui Fan$^a$,
Yun Soo Myung$^b$\footnote{ysmyung@inje.ac.kr;}, De-Cheng Zou$^{a,c}$\footnote{dczou@yzu.edu.cn;}\\
and Meng-Yun Lai$^{a}$\footnote{Corresponding author: mengyunlai@jxnu.edu.cn;}
}\\[8mm]

{${}^a$College of Physics and Communication Electronics, Jiangxi Normal University, Nanchang 330022, China\\[0pt] }
{${}^b$Institute of Basic Sciences and Department  of Computer Simulation,\\ Inje University, Gimhae 50834, Korea\\[0pt]}
{${}^c$Center for Gravitation and Cosmology, College of Physical Science and Technology, Yangzhou University, Yangzhou 225009, China\\[0pt]}
\end{center}

\vspace{2mm}
\vspace{2mm}

\begin{abstract}

We investigate  the tachyonic instability of  Kerr-Newman (KN) black hole with a rotation parameter $a$ in the Einstein-Chern-Simons-scalar theory coupled with a quadratic massive scalar field.
 This instability analysis corresponds to exploring the onset of spontaneous scalarization for KN black holes.
First, we find no $a$-bound for $\alpha<0$ case by considering  (1+1)-dimensional analytical  method.
A direct numerical  method is adopted  to explore (2+1)-dimensional time evolution of a massive scalar perturbation with   positive and negative $\alpha$ to obtain threshold curves numerically.
We obtain  threshold curves $\alpha_{\rm th}(a)$ of tachyonic instability for positive  $\alpha$ without any $a$-bounds.
We expect to find the same threshold curves $\alpha_{\rm th}(a)$ of tachyonic instability  for negative $\alpha$ without any $a$-bound because its linearized scalar theory is invariant under the transformation of  $\alpha\to -\alpha $ and $\theta\to -\theta$.  In addition, it is found that  the scalar mass term suppresses tachyonic instability of KN black holes.
\end{abstract}
\vspace{5mm}

\newpage
\renewcommand{\thefootnote}{\arabic{footnote}}
\setcounter{footnote}{0}

\section{Introduction}

Recently, there was  a significant progress on obtaining black holes with scalar hair by making use of  spontaneous scalarization. This corresponds to an example for evasion of no-hair theorem proposed in~\cite{Carter:1971zc,Ruffini:1971bza,Bekenstein:1974sf,Bekenstein:1995un}. In this case, the tachyonic instability of bald  black holes represents  the onset of scalarized black holes when introducing scalar couplings to the source terms: the Gauss-Bonnet term  for Schwarzschild black hole~\cite{Doneva:2017bvd,Silva:2017uqg,Antoniou:2017acq} and  Kerr black hole~\cite{Cunha:2019dwb} or Maxwell term  for Reissner-Nordstr\"{o}m black hole (RN)~\cite{Herdeiro:2018wub,Promsiri:2023yda} and  Kerr-Newman (KN) black hole~\cite{Lai:2022ppn}.
Source terms might include  either geometric invariant sources of Ricci scalar ($R$), Gauss-Bonnet term ($R^2_{\rm GB}$), and Chern-Simons term (Pontryagin density: ${}^*RR$) or matter invariant source (Maxwell term: $F^2$).
Here, it is worth noting  that  $R$ leads to a conformal coupling, $R^2_{\rm GB}$ takes  complicated forms for RN  and KN black holes~\cite{Herdeiro:2021vjo},  ${}^*RR$  is zero for static  Schwarzschild and RN black holes, and  $F^2$  is zero for non-charged  Schwarzschild and Kerr black holes.

 The study on tachyonic instability of Kerr black hole in the Einstein-Chern-Simons-scalar theory with a quadratic scalar coupling  was firstly  considered for  a massless scalar propgation~\cite{Gao:2018acg,Myung:2020etf,Doneva:2021dcc}. It is interesting to note that  this theory differs from the dynamical Chern-Simons gravity in the sense that the former has a linear scalar coupling to Chern-Simons term~\cite{Smith:2007jm,Alexander:2009tp}, while the latter shows a quadratic scalar coupling to Chern-Simons term. Hence, the dynamical Chern-Simons gravity could not be a candidate for spontaneous scalarization.
 Because its linearized scalar equation is invariant under the transformation of $\alpha\to -\alpha $ and $\theta\to -\theta$, one has recovered  the same threshold curve $[\log\alpha(a)]$ which is the boundary between Kerr and scalarized Kerr black holes  for positive and negative $\alpha$ without any $a$-bounds.  Secondly, considering  a massive scalar coupling to Chern-Simoms term~\cite{Zhang:2021btn} has  led to that the scalar mass term suppresses the tachyonic instability of Kerr black hole.

 On the other hand, the tachyonic instability of KN black hole was investigated  in the Einstein-Maxwell-scalar (EMS) theory with positive and negative scalar coupling to Maxwell term.
 For the positive coupling $\alpha$, a 3D graph [$\log_{10}\alpha(a,Q)$] was found, which represents the onset surface for the spontaneous scalarization of the KN black hole without any $a$-bounds~\cite{Lai:2022ppn}. On the contrary, for  negative coupling $\alpha$, the threshold curves $\log[-\alpha(a)]$ with $Q=0.1,0.4,0.7$ for spontaneous scalarization which describe boundaries between blad KN and scalarized KN black holes with $a$-bound of $a\ge 0.4142 r_+$~\cite{Lai:2022spn}. This implies that the case of $\alpha>0$ shows different results from the $\alpha<0$ case in the EMS theory.

 In the present work, we wish to investigate  the tachyonic instability of  KN  black hole in the Einstein-Chern-Simons-scalar theory with a quadratic massive scalar coupling (ECSs theory).
 We expect to have the same threshold curve $\alpha_{\rm th}(a)$ of tachyonic instability for positive and negative $\alpha$ without any $a$-bounds because its linearized scalar theory is invariant under the transformation of  $\alpha\to -\alpha $ and $\theta\to -\theta$.

The organization of our work is as follows. In  section 2, we will derive the linearized scalar equation, which is essential for studying the tachyonic instability of KN black holes.
Analytic computation will be done to see whether any $a$-bounds for spin-induced KN black hole scalarization exist or not  for the case of  $\alpha<0$ in section 3.
We describe a direct  numerical method to explore (2+1)-dimensional time evolution  of a massive scalar perturbation in section 4. We briefly explain  numerical results in section 5 by displaying three relevant figures.
Finally, in section 6,  we conclude our result and discuss some relevant results.

\section{Linearized scalar equation}\label{2s}

Our action of the Einstein-Chern-Simons-scalar theory with a quadratic massive scalar coupling (ECSs theory) takes the  form
\begin{eqnarray}
S_{\rm ECSs}=\frac{1}{16\pi}\int d^4x{\sqrt{-g}\Big[R-F^2-\frac{1}{2}\partial_\mu\phi\partial^\mu\phi+f(\phi)^*RR-U(\phi)\Big]}, \label{action}
\end{eqnarray}
where $\phi$ is a scalar field and $F^2=F_{\mu\nu}F^{\mu\nu}$ is the Maxwell kinetic term. Moreover, the coupling function $f(\phi)$ is important to control a nonminimal coupling of scalar $\phi$ to the CS invariant
\begin{eqnarray}
    ^*RR= \frac{1}{2}\epsilon^{\alpha\beta\gamma\delta}R^\mu_{\nu\gamma\delta}R^\nu_{\mu\alpha\beta}.
\end{eqnarray}
For simplicity, we introduce a quadratic scalar coupling function $f(\phi)$ and a potential  $U(\phi)$ defined  as
\begin{eqnarray}
    f(\phi) = \alpha \phi^2, \quad U(\phi) = \frac{1}{2}m^2_\phi\phi^2,
\end{eqnarray}
where $f(\phi)$
satisfies
\begin{eqnarray}\label{fphi}
f(0)=0,\quad \frac{d f}{d\phi}(0)=0, \quad \frac{d^2f}{d\phi^2}(0)=2\alpha, \label{fphi2}
\end{eqnarray}
and $m_\phi$ denotes a mass of the scalar field.

Varying  action \eqref{action} with respect to the scalar field $\phi$ vector potential $A_\mu$, and metric tensor $g_{\mu\nu}$ leads to
\begin{eqnarray}
		\square\phi &+&\frac{d f(\phi)}{d\phi}~^*RR-U'(\phi)=0,\label{s-equa1}\\
        \nabla^\mu F_{\mu\nu}&=&0,\\
		R_{\mu\nu}&-&\frac{1}{2}g_{\mu\nu}R = T^{\rm CS}_{\mu\nu}+T^\phi_{\mu\nu}+T^A_{\mu\nu},\label{MetricEq}\\
		T^{\rm CS}_{\mu\nu}&=& -4 \nabla^\sigma f \epsilon_{\sigma \alpha \beta (\mu} \nabla^\beta R_{\nu)}^{~\alpha} - 4 \nabla^\alpha \nabla^\beta f ~^\ast R_{\alpha (\mu\nu) \beta},\nonumber\\
		T^\phi_{\mu\nu}&=&\frac{1}{2}\Big[\partial_\mu\phi\partial_\nu\phi-\frac{1}{2} g_{\mu\nu}\partial^\rho\phi\partial_\rho \phi-g_{\mu\nu}U(\phi)\Big],\nonumber\\
         T^A_{\mu\nu}&=&2F_{\mu\rho}F_{\nu}~^\rho-\frac{1}{2}F^2g_{\mu\nu}.\nonumber
	\end{eqnarray}
Without scalar hair ($\phi=0$), this theory admits the axisymmetric Kerr-Newman (KN) black hole solution expressed in terms of the Boyer-Lindquist coordinates as
\begin{eqnarray}
ds^2_{\rm KN} &\equiv& \bar{g}_{\mu\nu}dx^\mu dx^\nu=-\frac{\Delta-a^2\sin^2\theta}{\rho^2}dt^2
-\frac{2a\sin^2\theta(r^2+a^2-\Delta)}{\rho^2}dt d\varphi\nonumber\\
&&+\frac{[(r^2+a^2)^2-\Delta a^2 \sin^2\theta]\sin^2\theta}{\rho^2} d\varphi^2+ \frac{\rho^2}{\Delta}dr^2 +\rho^2 d\theta^2 \label{KN-sol}
\end{eqnarray}
with two parameters
\begin{eqnarray}
 \Delta= r^2-2Mr+a^2+Q^2\quad {\rm and} \quad \rho^2=r^2+a^2\cos^2\theta.\nonumber
\end{eqnarray}
Then, the outer and inner horizons could be  found  by imposing  $\Delta=(r-r_+)(r-r_-)=0$ as
\begin{eqnarray}
r_{\pm}=M\pm \sqrt{M^2-Q^2-a^2}.
\end{eqnarray}
At this stage,  one requires an  existence condition for the outer horizon
\begin{equation}
M^2-Q^2\ge a^2. \label{ex-con}
\end{equation}
Curiously, the CS  term in the KN background takes  a complicated form
\begin{eqnarray}
^*\bar{R}\bar{R}&=&\frac{96M^2 r a\cos\theta}{(r^2+a^2\cos^2\theta)^6}\left[(3r^2-a^2\cos^2\theta)-\frac{2Q^2}{M}r\right]\nonumber\\
&&\times\left[(r^2-3a^2\cos^2\theta)-\frac{Q^2}{Mr}(r^2-a^2\cos^2\theta)\right].\label{effmass1}
\end{eqnarray}
We observe that in the limit of  $Q\to0$, Eq.(\ref{effmass1}) reduces to that for Kerr black hole found in the ECSs theory~\cite{Myung:2020etf}.
On the other hand,  one finds that $^*\bar{R}\bar{R}\to 0$  in the limit of $a\to0$, implying  that there is no static limit.
Furthermore, we confirm that the CS term becomes parity odd under the transformation of $\theta\to \pi-\theta$ because of ``$\cos \theta$" in the front.

To perform the tachyonic instability of KN black holes, we need to  obtain  the linearized scalar equation by linearizing Eq.(\ref{s-equa1}) as
\begin{eqnarray}
\left(\bar{{\square}}-\mu^2_{\rm s}\right)\delta\phi=0,\quad \mu^2_{\rm s}=m_\phi^2-2\alpha^*\bar{R}\bar{R} \label{per-eq}.
\end{eqnarray}
A linearized scalar field  acquires an effective mass square $\mu^2_{\rm s}=m_\phi^2 -2\alpha ~^\ast R R$, which depends on $r$ and $\theta$ and allows to have negative values near the outer horizon (see Fig. 1).
It may lead  to tachyonic instability. Furthermore, It is important to note that the linearized scalar equation (\ref{per-eq}) is invariant under the transformation
	\begin{eqnarray}
    \alpha \rightarrow -\alpha,\quad \theta \rightarrow \pi-\theta. \label{parity}
	\end{eqnarray}	
Hence, one expects that the case of $\alpha<0$  is  the same as the case with $\alpha>0$ in the KN background (see Fig. 1), as has been confirmed numerically and analytically in the Kerr black hole background~\cite{Myung:2020etf}.
In the section 3, therefore, we wish to carry out analytic computation for $\alpha<0$ case to find that there is ``no $a$-bound in the KN background.
In the  section 4,  we will  consider the $\alpha>0$ case  and study carefully the time evolution of a massive scalar  perturbation to find  tachyonic instability.
\begin{figure*}[t!]
   \centering
  \includegraphics{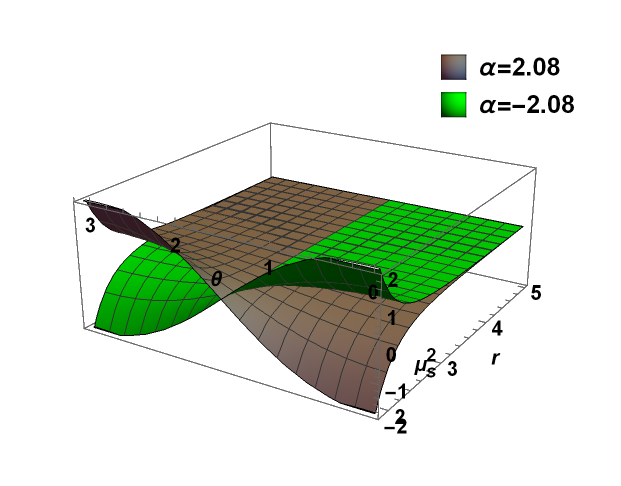}
\caption{3D pictures of effective scalar mass $\mu^2_s(r,\theta)$  for $m_\phi=0.25$, $Q=0.4$ and
$a=0.2$ with two opposite  couplings $\alpha=\pm2.08$. Here, $r\in[r_+=1.89,5]$ and $\theta\in[0,\pi]$. They  show odd parity under the transformation of $\alpha\to -\alpha$. }\label{fig1}
\end{figure*}	
\section{Analytic results for $\alpha<0$ case}\label{3s}

Before we proceed, we would like to  mention that the tacyonic instability condition  of Kerr black holes was  found as $a$-bound of $a \ge 0.5M$ in the Einstein-Gauss-Bonnet-scalar (EGBs) theory with  $\alpha<0$ by solving the (1+1)-dimensional linearized scalar equation~\cite{Dima:2020yac}. This was confirmed analytically in~\cite{Hod:2020jjy} and numerically in~\cite{Doneva:2020nbb}. This is so-called the spin-induced black hole scalarization in EGBs theory with negative coupling  because this scalarization has never found in the static black holes.
Hence, it is very  curious to check  whether such a bound exists or not in the KN background for the ECSs theory  with $\alpha<0$.

In this section, we will check  that there is ``no $a$-bound" analytically in the (1+1)-dimensional linearized scalar equation.
Hereafter, we set the mass of KN black hole to be $M=1$ for simplicity.
As shown in Ref. \cite{Dima:2020rzg}, we can also introduce two coordinates transformations
\begin{eqnarray}
d\varphi^*=d\varphi +\frac{a}{\Delta} dr,\quad dx=\frac{r^2+a^2}{\Delta} dr.
\end{eqnarray}
Now, we can express  Eq.(\ref{per-eq}) in terms of the Boyer-Lindquist coordinates
\begin{eqnarray}
  &&\left[ (r^2+a^2)^2-
      \Delta a^2\sin^2\theta\right]\partial^2_t \delta\phi
    -(r^2+a^2)^2\partial_x^2\delta\phi
    -2r\Delta\partial_x\delta\phi\nonumber\\
    &&+2a\left(2r-Q^2\right)\partial_t\partial_{\varphi^*}\delta\phi
    -2a(r^2+a^2)\partial_x\partial_{\varphi^*}\delta\phi
   \nonumber\\
    &&-\Delta\left[\frac{1}{\sin\theta}\partial_\theta(\sin\theta\partial_\theta\delta\phi)
    +\frac{1}{\sin^2\theta}\partial^2_{\varphi^*}\delta\phi\right]\nonumber\\
    &&+\Delta\left(r^2+a^2\cos^2\theta\right)\mu_{\rm s}^2\delta\phi=0. \label{perturbedEq2-1}
\end{eqnarray}
At this stage, we wish to  introduce a projection of linearized scalar equation (\ref{perturbedEq2-1}) onto a basis of spherical harmonics. Let us consider  a decomposition of $\delta \phi$  in terms of  spherical harmonics $Y_{lm}(\theta,\varphi)$
\begin{eqnarray}
 \delta \phi(t,r,\theta,\varphi) =\sum_{m}\sum_{l=|m|}^{\infty} \frac{\psi_{l m}(t,r)}{r} Y_{l m}(\theta,\varphi).
\end{eqnarray}
We plug this decomposition into Eq.(\ref{perturbedEq2-1}) and obtain a coupled $(1+1)$-dimensional evolution equation for $\psi_{l m}\equiv\int r\delta\phi Y^*_{l m}d\Omega$
\begin{eqnarray}
&&\Big[(r^2+a^2)^2-a^2\Delta(1-c^m_{ll})\Big]\ddot{\psi}_{l m}+
a^2\Delta\left(c^m_{l,l+2}\ddot{\psi}_{l+2,m}+c^m_{l,l-2}\ddot{\psi}_{l-2,m}\right)\nonumber\\
&&+2i a m \left(2 r-Q^2\right)\dot{\psi}_{l m}-(r^2+a^2)^2\psi''_{l m}-
\Big[2i a m(r^2+a^2)-2a^2\Delta/r\Big]\psi'_{l m}\nonumber\\
&&\Delta\Big[l(l+1)+\frac{2M}{r}-\frac{2a^2+2Q^2}{r^2}+\frac{2i a m}{r}\Big]\psi_{l m}\nonumber\\
&&+\Delta\sum_{j m'}<l m|\mu^2_{\rm s}(r^2+a^2\cos^2\theta)|j m'>\psi_{j m'}=0.
\end{eqnarray}
Here, the overdot ($\dot{}$) represents a derivative with respect to time $t$, while the prime (${}'$) denotes a derivative with respect to  $x$.
Different $m$ modes decouple from one another, whereas the decomposition in spherical harmonics generates couplings for each $l$ mode to the $l\pm2$ modes  because of the axisymmetric nature of the KN spacetime.
On the threshold  of tachyonic instability, the last term of
$\Delta\sum_{jm'}< l m | \mu^2_{\rm s}(r^2+a^2\cos^2\theta)| j m'>\psi_{j m'}$ can be replaced by a single term as
\begin{equation}
\Delta< l_1 m=0 | \mu^2_{\rm s}(r^2+a^2\cos^2\theta)| l_2 m=0>\psi_{l_2 m}
\end{equation}
at asymptotically late times.
We note that the onset of spontaneous scalarization is related to  an effective  binding potential well appearing near the outer horizon whose two turning points of $r_{\rm in}$ and $r_{\rm out}$ are determined  by the relation of $r_{\rm out}\ge r_{\rm in}=r_+$.  A critical KN black hole with $a=a_{\rm c}$  is defined by  a boundary between KN  and scalarized KN black holes  existing  in the limit of $\alpha \to -\infty$. In this sense,
the mass term $m^2_\phi$ in $\mu^2_s$ does not play any roles.  This is usually  characterized by  the presence of a degenerate  binding potential well whose two turning points
merge at the outer horizon ($r_{\rm out}= r_{\rm in}=r_+$) as
\begin{eqnarray}
< l_1 m=0 | \mu^2_{\rm s}(r^2+a^2\cos^2\theta)| l_2 m=0>|_{r=r_+,~a=a_{\rm c}}=0
\end{eqnarray}
in the limit of $\alpha \to -\infty$.
Here, the critical rotation parameter $a_{\rm c}$ is determined simply by the resonance condition
\begin{eqnarray}
\int_0^\pi\frac{a_{\rm c}r_+ \cos \theta A(r_+,a_{\rm c}) }{(r_+^2+a^2_{\rm c}\cos^2\theta)^5}Y_{l_10}Y_{l_20} \sin\theta d\theta =0,\label{res-con}
\end{eqnarray}
where
\begin{eqnarray}
A(r_+,a_{\rm c})&=&\Big[(3r^2_+-a^2_{\rm c}\cos^2\theta)-2Q^2r_+\Big]  \nonumber \\
&\times& \Big[(r^2_+-3a^2_{\rm c}\cos^2\theta)-\frac{Q^2}{r_+}(r^2_+-a^2_{\rm c}\cos^2\theta)\Big].\label{effmass}
\end{eqnarray}
In order to solve Eq.~\eqref{res-con} for $a_{\rm c}$ analytically, it would be better to introduce three new variables
\begin{eqnarray}
\hat{a}\equiv\frac{a_{\rm c}}{r_+}, \qquad
X=\hat{a} \cos \theta, \qquad q=\frac{Q}{\sqrt{r_+}}.
\end{eqnarray}

Then, Eq.(\ref{res-con}) takes the form
\begin{eqnarray} \label{con-im}
\int^{\hat{a}}_{-\hat{a}} \frac{X(3X^2-2q^2-1)(X^2-3-q^2X^2+q^2)}{(1+X^2)^5}  Y_{l_10}(X/\hat{a})Y_{l_20}(X/\hat{a})dX=0.
\end{eqnarray}
In the asymptotic limit of $l_1=l_2 =l\to \infty$, one finds $Y^2_{l0}\to \delta(\theta)$ around the poles of  $\theta=0,\pi(X=\hat{a},-\hat{a})$.
In case of $X\to\hat{a}$, Eq.~\eqref{con-im} leads to a condition of the  resonance for $\hat{a}$ as
\begin{eqnarray}
\hat{a}(3\hat{a}^2-2q^2-1)(\hat{a}^2(1-q^2)-3+q^2)=0\label{hata}
\end{eqnarray}
which implies three solutions as
\begin{equation}
\hat{a}=0,\quad \sqrt{\frac{1}{3}+\frac{2q^2}{3}},\quad \sqrt{\frac{3-q^2}{1-q^2}}.
\end{equation}
Here, we seek for  the smallest non-zero value of the black hole spin parameter $\hat{a}$ which may allow the existence of  nonminimally coupled scalar clouds
for the critical  rotation parameter.
Taking $Q=0.4$, for example, the middle term of $\hat{a}=a_{\rm c}/r_+=\sqrt{1/3+0.32/(3r_+)}$  is used  to obtain
the critical rotation parameter \begin{eqnarray}
a_{\rm c}=0.8551.
\end{eqnarray}
In this case, it may imply the $a$-bound
\begin{eqnarray}
a\ge a_{\rm c} (=0.8551) \label{abound}
\end{eqnarray}
in the KN black hole background.
On the other hand, $a\ge a_{\rm c}(=0)$ obtained from $\hat{a}=0$   confirms ``no $a$-bound" and   Fig. 3 indicates  $\alpha$-dependent bounds  only.
It seems that Eq.(\ref{abound})  represents  an enhanced (instability) bound for scalarization in the ECSs theory with negative coupling parameter.
Unfortunately, there is no way to prove this enhanced $a$-bound numerically  in the KN background for the ECSs theory  with $\alpha<0$.
In the case of Kerr spacetime ($q=0$)~\cite{Myung:2020etf}, one has  found  $a\ge a_c$ with $a_{\rm c}=0$ and 0.866 where the former represents ``no $a$-bound'', while the latter denotes an unproven $a$-bound for scalarization
in the ECSs theory with negative coupling parameter. Hence,  Eq.(\ref{abound}) corresponds to the latter bound.
In the case of a scalar coupled to Maxwell term in the KN spacetime~\cite{Lai:2022spn}, however, one has found that ``$a\ge a_{\rm c}(=0.6722)$'' corresponds to a proven $a$-bound in the EMS theory with negative coupling parameter. This is spin-induced black hole scalarization.
Therefore,  the slowly rotating KN black hole with $a<0.6722$ could not develop  tachyonic instability and thus, cannot have scalarized KN black holes in the EMS theory.

\section{Numerical Method}\label{4s}
In this section, we wish to describe briefly a numerical method to study  tachyonic instability of KN black hole in the ECSs theory with positive scalar coupling.

Before  proceed, we would like to mention that it is not easy to solve the partial differential equation \eqref{perturbedEq2-1} directly.
In our previous work \cite{Lai:2022ppn},
we adopted the $(2+1)$-dimensional hyperboloidal foliation method to solve the linearized scalar equation
numerically with positive  coupling parameter.
However, the hyperboloidal foliation method is not suitable for a massive scalar field. Hence, in the present work we adopt a more direct numerical method to explore the $(2+1)$ time evolution of the scalar perturbation. After separating out the azimuthal dependence
\begin{eqnarray}
  \delta \phi (t,x,\theta,\varphi^*) &=& \sum_m \delta\phi(t,x,\theta)e^{im\varphi^*},
\end{eqnarray}
the linearized scalar equation (\ref{perturbedEq2-1}) reduces to
\begin{eqnarray}
  &&\left[ (r^2+a^2)^2-\Delta a^2\sin^2\theta\right]\partial^2_t\delta\phi -(r^2+a^2)^2\partial^2_x\delta\phi
  -\Delta\partial^2_\theta\delta\phi\nonumber\\
  &&+ 2ima{(2Mr-Q^2)}\partial_t\delta\phi-2\left[r\Delta+ima(r^2+a^2)\right]\partial_x\delta\phi-\Delta\cot{\theta}\partial_\theta\delta\phi \nonumber\\
  &&+  \Delta\Big[(r^2+a^2\cos^2\theta)\mu_{s}^2+\frac{m^2}{\sin^2\theta}\Big]\delta\phi= 0, \label{mscalar-eq2}
\end{eqnarray}
where $m$ is the azimuthal mode number. For rotating solutions, it is not possible to rigorously separate the $l$-dependence since different $l$ modes couple to each other. Therefore, we have to solve Eq.~\eqref{mscalar-eq2} numerically. For this purpose, it is convenient to rewrite this equation as the $(2+1)$-dimensional Teukolsky equation~\cite{Lai:2022ppn}
\begin{eqnarray}\label{perturbedEq8-2}
\partial^2_t\delta\phi + A^{xx}\partial^2_x\delta\phi
  +A^{\theta\theta}\partial^2_\theta\delta\phi  +B^{t}\partial_t\delta\phi + B^{x}\partial_x\delta\phi+ B^{\theta}\partial_\theta\delta\phi +C \delta\phi= 0,
\end{eqnarray}
where the coefficients are
\begin{eqnarray}\label{coeff8-0}
  A^{tt} &=& \left[ (r^2+a^2)^2-\Delta a^2\sin^2\theta\right] , \nonumber\\
  A^{xx} &=&-\frac{(r^2+a^2)^2}{A^{tt}},
  \nonumber\\
  A^{\theta\theta} &=&-\frac{\Delta}{A^{tt}},  \nonumber\\
  B^{t} &=& \frac{ 2ima{(2Mr-Q^2)}}{A^{tt}},
  \nonumber\\
  B^{x} &=& -\frac{ 2r\Delta+2ima(r^2+a^2)}{A^{tt}},
  \nonumber\\
  B^{\theta} &=&-\frac{ \Delta\cot{\theta}}{A^{tt}},
  \nonumber\\
  C &=& \frac{\Delta}{A^{tt}}\left[(r^2+a^2\cos^2\theta)\mu_{s}^2+\frac{m^2}{\sin^2\theta}\right].
\end{eqnarray}
For $Q=0$, Eq.~\eqref{perturbedEq8-2} reduces exactly to Eq. (12) in Ref.~\cite{Zhang:2021btn}.

Introducing the following auxiliary fields and dividing them into real and imaginary parts
\begin{eqnarray}
  \Phi &\equiv& \delta\phi, \nonumber\\
  \Psi &\equiv& \partial_x \Phi, \nonumber\\
  \Pi  &\equiv& \partial_t \Phi, \label{aux-f}
\end{eqnarray}
Eq.~\eqref{perturbedEq8-2} can be rewritten as two coupled equations
\begin{eqnarray}
 \partial_t\Pi_R=&-&\Big( A^{xx}\partial_x\Psi_R
  +A^{\theta\theta}\partial^2_\theta\Phi_R \nonumber \\
&-&B^t_I\Pi_I + B^x_R\Psi_R- B^x_I\Psi_I+B^\theta\partial_\theta\Phi_R+C\Phi_R\Big),
  \label{eqsv-1}\\
 \partial_t\Pi_I=&-&\Big(  A^{xx}\partial_x\Psi_I
  +A^{\theta\theta}\partial^2_\theta\Phi_I \nonumber \\
&+&B^t_I\Pi_R + B^x_I\Psi_R+ B^x_R\Psi_I +B^\theta\partial_\theta\Phi_I+C\Phi_I\Big),\label{eqsv-2}
\end{eqnarray}
where the subscripts $R$ and $I$ denotes the real and image parts of the auxiliary fields respectively. For simplicity, the above equations can be rewritten compactly as
\begin{eqnarray}
  \partial_t u &=& (G\partial_x+Y)u, \label{comp-eq}
\end{eqnarray}
where $u= (\Phi_R,\Phi_I,\Psi_R,\Psi_I,\Pi_R,\Pi_I)^T$ and
\begin{eqnarray}\label{perturbedEq3}
  G &=&  \left(
           \begin{array}{cccccc}
             0 & 0 & 0 & 0 & 0 & 0 \\
             0  & 0 & 0 & 0 & 0 & 0 \\
             0 & 0 & 0 & 0 & 1 & 0 \\
             0 & 0 & 0 & 0 & 0 & 1 \\
             0 & 0 & G_{53} & 0 & 0 & 0 \\
             0 & 0 & 0 & G_{64} & 0 & 0 \\
           \end{array}
         \right),
  \label{G-mat}\\
   Y &=&  \left(
           \begin{array}{cccccc}
             0 & 0 & 0 & 0 & 1 & 0 \\
             0  & 0 & 0 & 0 & 0 & 1 \\
             0 & 0 & 0 & 0 & 0 & 0 \\
             0 & 0 & 0 & 0 & 0 & 0 \\
             Y_{51} & 0 & Y_{53} & Y_{54} & 0 & Y_{56} \\
             0 & Y_{62}  & Y_{63}  & Y_{64}  & Y_{56}  & 0 \\
           \end{array}
         \right) \label{Y-mat}
\end{eqnarray}
with matrix elements
\begin{eqnarray}
  G_{53} &=& G_{64}=-A^{xx}, \nonumber \\
  Y_{51} &=& Y_{62}= -(A^{\theta\theta}\partial_\theta^2+B^\theta\partial_\theta+C),
  \nonumber\\
  Y_{53} &=& Y_{64}=B^x_R, \nonumber\\
  Y_{54} &=& -Y_{63}=-B^x_I,
  \nonumber\\
  Y_{56}&=&Y_{65}=B^t_I. \label{mat-coeff}
\end{eqnarray}
Now, the linearized scalar equation \eqref{comp-eq} becomes a first-order equation  in time. Therefore, the time evolution can be rapidly carried out by adopting the fourth-order Runger-Kutta integrator, while  derivatives in $x$ and $\theta$ directions are approximated by making use of the high-order finite difference formulae.

In the above, we introduced the tortoise coordinate $x\in(-\infty,\infty)$. Two boundary conditions are set as ingoing (outgoing) boundary conditions at $x=-\infty(x=\infty)$. Numerically, the infinities must be truncated at finite values, which will inevitably cause spurious reflections from the outer boundary. Fortunately, we can protect the scalar signal from the pollution of the spurious reflections for a long time by pushing the numerical outer boundary sufficiently far away. By doing so, we increase the computation time in exchange for satisfactorily precise scalar signals. At the poles of $\theta=0,\pi$, we simply impose the boundary conditions of $\Phi|_{\theta=0,\pi}=0$ for $m\not=0$, whereas $\partial \Phi|_{\theta=0,\pi}=0$ for $m=0$.

To implement the fourth-order Runger-Kutta integrator, we impose a spherically harmonic Gaussian
distribution as an initial condition
\begin{eqnarray}
\Phi(t=0,x,\theta)\sim Y_{lm}(\theta,0)e^{-\frac{(x-x_c)^2}{2\sigma^2}}, \label{ini-S}
\end{eqnarray}
where $x_c$ represents a location of the center and $\sigma$ denotes  a width of the distribution. It is worth to point out that although  there is one initial mode with a specified $l$ only,
other $l$ modes with the same index $m$ will be activated during the evolution. It is necessary to consider the influence of $l$ and $m$ of the initial perturbation. However, we note that the $l=m$ mode will have a dominant contribution at late times. To our knowledge,  similar phenomena have  occurred in other theories. Hence,  we choose the initial perturbations with $l=m=0$ only for simplicity.
The above numerical scheme has been successfully implemented in the literatures~\cite{Zhang:2021btn,Lai:2022spn,Doneva:2020nbb}.
\begin{figure*}[t!]
   \centering
  \includegraphics{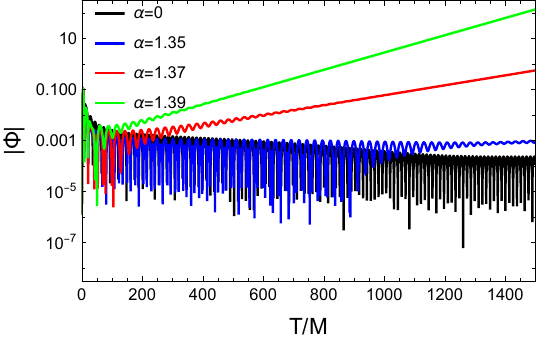}
\caption{Time-domain profiles of the linearized scalar field $|\Phi|$ for $m_\phi=0.25$, $Q=0.4$ and
$a=0.3$ with four different coupling $\alpha$. Here, $\alpha=1.35$ represents a threshold value $\alpha_{\rm th}$. $\alpha=1.37,~1.39$ are unstable modes and $\alpha=0$ is a stable mode. }\label{fig2}
\end{figure*}
\begin{figure*}[t!]
   \centering
  \includegraphics{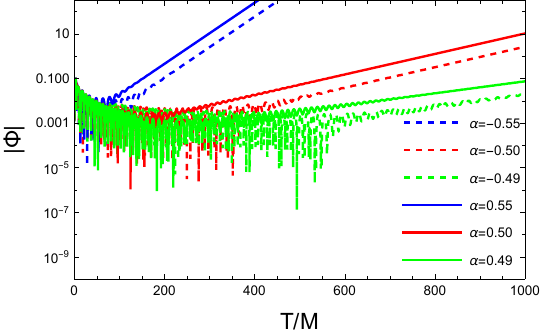}
\caption{Similar time-domain profiles of the linearized scalar field $|\Phi|$ for $m_\phi=0.5$, $Q=0.4$ and
$a=0.8$ with six different  couplings $\alpha=\mp0.55,\mp0.50,$ and $\mp0.49$. Each of three opposite pairs shows similar unstable nature. }\label{fig3}
\end{figure*}
\section{Numerical Results}\label{5s}
In this section,  we describe  schematic  pictures on the influences of scalar coupling parameter $\alpha$ and  scalar  mass $m_\phi$ on scalar wave dynamics  by  studying  the (2+1)-dimensional time evolution of a massive scalar  perturbation in  the ECSs theory.
Considering different coupling parameter $\alpha$, the time domain profiles of the $l=m=0$-scalar mode $|\Phi|$ are shown in Fig. 2.

A case of $\alpha_{\rm th}=1.35$  represents the threshold (marginal) evolution of tachyonic instability for $m_\phi=0.25,~Q=0.4$, and $a=0.3$. From this figure, it is easily known that tachyonic instability is  triggered once the coupling constant $\alpha$ exceeds a threshold value $\alpha_{\rm th}$. Importantly, Fig. 3 indicates  time-domain profiles of the $l=m=0$-scalar mode with $m_\phi=0.5,~Q=0.4$, and $a=0.8$ for six different couplings $\alpha=\mp0.55,\mp0.50,$ and $\mp0.49$.

\begin{figure*}[t!]
   \centering
  \includegraphics[width=0.5\textwidth]{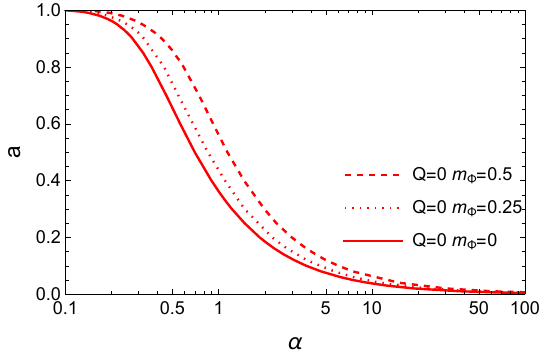}
  \hfill%
  \includegraphics[width=0.5\textwidth]{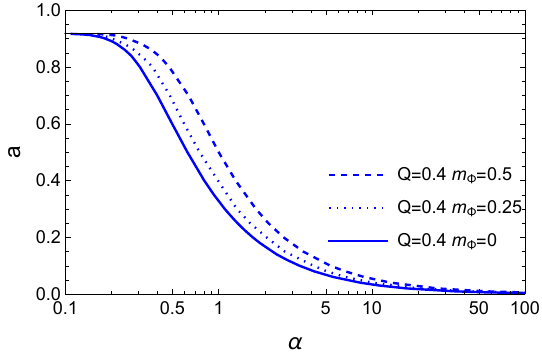}
  \hfill%
  \includegraphics[width=0.5\textwidth]{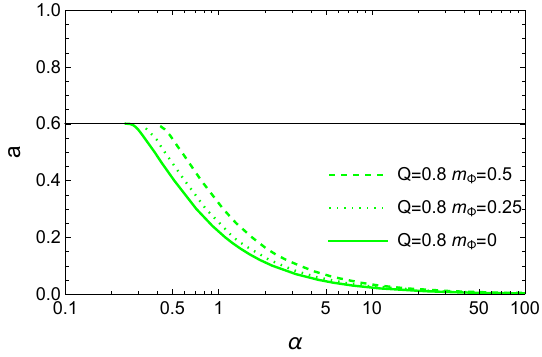}
\caption{Threshold curves $\alpha_{\rm th}(a)$ for different charge $Q=0,~0.4,~0.8$ with different scalar mass $m_\phi=0,~0.25,~0.5$.  All colored curves denote the boundaries between stable (lower) and unstable (upper) region. Black solid lines
represent the existence condition ($M^2-Q^2\ge a^2$) of the outer horizon.}\label{fig4}
\end{figure*}

As was pointed out, one expects that  the case of $\alpha<0$ is nearly  the same as the case of $\alpha>0$ in the KN background because of the symmetry in Eq.(\ref{parity}).
We observe from  Fig. 3 that  each  pair of opposite  signs for  $\alpha$ shows  similar time-domain profiles $|\Phi|$.
Hence, we consider the case of $\alpha>0$ for simplicity.

A more clear picture for influences of three parameters $(a, \alpha, m_\Phi)$ on the onset of  tachyonic instability is depicted  in Fig. 4.
For $m_\phi=0$, the scalar field becomes massless and its threshold curves are located in the lowest for a given $\alpha$. This implies that the presence of a positive  scalar mass term suppresses  or even  quench the tachyonic instability~\cite{Zhang:2021btn,Zou:2019bpt}. Considering the existence condition of the outer horizon Eq.(\ref{ex-con}), one could determine  the upper bound of $a$.
For $Q=0,~0.4,$ and 0.8, the maximum values (upper bounds) of $a$ are given by 1, 0.917, and 0.6, respectively.  For $\alpha<0$, we expect to obtain the same graphs as shown  in Fig. 4
because of the symmetry under the transformation Eq.(\ref{parity}). Hence, we do not wish to  display the corresponding pictures for $\alpha<0$.

\section{Conclusions and discussions}\label{6s}

We have  investigated  the tachyonic instability of  Kerr-Newman (KN) black hole in the Einstein-Chern-Simons-scalar theory  with a quadratic massive scalar coupling (ECSs theory).
This analysis  corresponds to exploring the onset of spontaneous scalarization for KN black holes.
For this purpose, we confirm that there is  no $a$-bound for $\alpha<0$ case by adopting (1+1)-dimensional analytical  method proposed by Hod~\cite{Hod:2020jjy}.
This case contrasts to the KN black hole found in the Einstein-Maxwell-scalar theory where an $a$-bound was known to be $a\ge a_c(=0.6722)$~\cite{Lai:2022spn}, whereas this is consistent with Kerr black hole found in the ECSs theory~\cite{Myung:2020etf}.

We have performed the (2+1)-dimensional time evolution of a scalar perturbation with   positive  $\alpha$ to obtain threshold curves numerically by choosing a direct numerical method.
We have obtained  threshold curves $\alpha_{\rm th}(a)$ of tachyonic instability for positive $\alpha$ with three different $Q=0,0.4,0.8$ without any $a$-bounds.
We expect to have the same curves $\alpha_{\rm th}(a)$ of tachyonic instability even for negative  $\alpha$  with three different $Q=0,0.4,0.8$ without any $a$-bounds because its linearized scalar theory is invariant under the transformation of  $\alpha\to -\alpha $ and $\theta\to -\theta$.  In addition, it is shown  that  the scalar mass term suppresses tachyonic instability of KN black holes.

 \vspace{1cm}

{\bf Acknowledgments}

M. Y. L is supported by the National Natural Science Foundation of China with Grant No. 12305064 and Jiangxi Provincial Natural Science Foundation with Grant No. 20224BAB211020. D. C. Z is supported by National Natural Science Foundation of China (NSFC) (Grant No. 12365009) and Jiangxi Provincial Natural Science Foundation (No. 20232BAB201039).


\vspace{1cm}





\newpage

\begin{thebibliography}{99}

\bibitem{Carter:1971zc}
B.~Carter,
Phys. Rev. Lett. \textbf{26}, 331-333 (1971)
doi:10.1103/PhysRevLett.26.331

\bibitem{Ruffini:1971bza}
R.~Ruffini and J.~A.~Wheeler,
Phys. Today \textbf{24}, no.1, 30 (1971)
doi:10.1063/1.3022513

\bibitem{Bekenstein:1974sf}
J.~D.~Bekenstein,
Annals Phys. \textbf{82}, 535-547 (1974)
doi:10.1016/0003-4916(74)90124-9

\bibitem{Bekenstein:1995un}
J.~D.~Bekenstein,
Phys. Rev. D \textbf{51}, no.12, R6608 (1995)
doi:10.1103/PhysRevD.51.R6608


\bibitem{Doneva:2017bvd}
D.~D.~Doneva and S.~S.~Yazadjiev,
Phys. Rev. Lett. \textbf{120}, no.13, 131103 (2018)
doi:10.1103/PhysRevLett.120.131103
[arXiv:1711.01187 [gr-qc]].


\bibitem{Silva:2017uqg}
H.~O.~Silva, J.~Sakstein, L.~Gualtieri, T.~P.~Sotiriou and E.~Berti,
Phys. Rev. Lett. \textbf{120}, no.13, 131104 (2018)
doi:10.1103/PhysRevLett.120.131104
[arXiv:1711.02080 [gr-qc]].

\bibitem{Antoniou:2017acq}
G.~Antoniou, A.~Bakopoulos and P.~Kanti,
Phys. Rev. Lett. \textbf{120}, no.13, 131102 (2018)
doi:10.1103/PhysRevLett.120.131102
[arXiv:1711.03390 [hep-th]].

\bibitem{Cunha:2019dwb}
P.~V.~P.~Cunha, C.~A.~R.~Herdeiro and E.~Radu,
Phys. Rev. Lett. \textbf{123}, no.1, 011101 (2019)
doi:10.1103/PhysRevLett.123.011101
[arXiv:1904.09997 [gr-qc]].

\bibitem{Herdeiro:2018wub}
C.~A.~R.~Herdeiro, E.~Radu, N.~Sanchis-Gual and J.~A.~Font,
Phys. Rev. Lett. \textbf{121}, no.10, 101102 (2018)
doi:10.1103/PhysRevLett.121.101102
[arXiv:1806.05190 [gr-qc]].
\bibitem{Promsiri:2023yda}
C.~Promsiri, T.~Tangphati, E.~Hirunsirisawat and S.~Ponglertsakul,
Phys. Rev. D \textbf{108} (2023) no.2, 024015
doi:10.1103/PhysRevD.108.024015
[arXiv:2302.04654 [gr-qc]].
\bibitem{Lai:2022ppn}
M.~Y.~Lai, Y.~S.~Myung, R.~H.~Yue and D.~C.~Zou,
Phys. Rev. D \textbf{106}, no.8, 084043 (2022)
doi:10.1103/PhysRevD.106.084043
[arXiv:2208.11849 [gr-qc]].

\bibitem{Herdeiro:2021vjo}
C.~A.~R.~Herdeiro, A.~M.~Pombo and E.~Radu,
Universe \textbf{7}, no.12, 483 (2021)
doi:10.3390/universe7120483
[arXiv:2111.06442 [gr-qc]].


\bibitem{Gao:2018acg}
Y.~X.~Gao, Y.~Huang and D.~J.~Liu,
Phys. Rev. D \textbf{99}, no.4, 044020 (2019)
doi:10.1103/PhysRevD.99.044020
[arXiv:1808.01433 [gr-qc]].
	

	
\bibitem{Myung:2020etf}
Y.~S.~Myung and D.~C.~Zou,
Phys. Lett. B \textbf{814}, 136081 (2021)
doi:10.1016/j.physletb.2021.136081
[arXiv:2012.02375 [gr-qc]].

\bibitem{Doneva:2021dcc}
D.~D.~Doneva and S.~S.~Yazadjiev,
Phys. Rev. D \textbf{103}, no.8, 083007 (2021)
doi:10.1103/PhysRevD.103.083007
[arXiv:2102.03940 [gr-qc]].


\bibitem{Smith:2007jm}
T.~L.~Smith, A.~L.~Erickcek, R.~R.~Caldwell and M.~Kamionkowski,
Phys. Rev. D \textbf{77}, 024015 (2008)
doi:10.1103/PhysRevD.77.024015
[arXiv:0708.0001 [astro-ph]].

\bibitem{Alexander:2009tp}
S.~Alexander and N.~Yunes,
Phys. Rept. \textbf{480}, 1-55 (2009)
doi:10.1016/j.physrep.2009.07.002
[arXiv:0907.2562 [hep-th]].

\bibitem{Zhang:2021btn}
S.~J.~Zhang,
Eur. Phys. J. C \textbf{81}, no.5, 441 (2021)
doi:10.1140/epjc/s10052-021-09249-8
[arXiv:2102.10479 [gr-qc]].

\bibitem{Lai:2022spn}
M.~Y.~Lai, Y.~S.~Myung, R.~H.~Yue and D.~C.~Zou,
Phys. Rev. D \textbf{106}, no.4, 044045 (2022)
doi:10.1103/PhysRevD.106.044045
[arXiv:2206.11587 [gr-qc]].

\bibitem{Dima:2020yac}
A.~Dima, E.~Barausse, N.~Franchini and T.~P.~Sotiriou,
Phys. Rev. Lett. \textbf{125}, no.23, 231101 (2020)
doi:10.1103/PhysRevLett.125.231101
[arXiv:2006.03095 [gr-qc]].

\bibitem{Hod:2020jjy}
S.~Hod,
Phys. Rev. D \textbf{102}, no.8, 084060 (2020)
doi:10.1103/PhysRevD.102.084060
[arXiv:2006.09399 [gr-qc]].

\bibitem{Doneva:2020nbb}
  D.~D.~Doneva, L.~G.~Collodel, C.~J.~Krüger and S.~S.~Yazadjiev,
  Phys.\ Rev.\ D {\bf 102}, no. 10, 104027 (2020)
  doi:10.1103/PhysRevD.102.104027
  [arXiv:2008.07391 [gr-qc]].

\bibitem{Dima:2020rzg}
A.~Dima and E.~Barausse,
Class. Quant. Grav. \textbf{37}, no.17, 175006 (2020)
doi:10.1088/1361-6382/ab9ce0
[arXiv:2001.11484 [gr-qc]].





\bibitem{Zou:2019bpt}
D.~C.~Zou and Y.~S.~Myung,
Phys. Rev. D \textbf{100}, no.12, 124055 (2019)
doi:10.1103/PhysRevD.100.124055
[arXiv:1909.11859 [gr-qc]].
\end{thebibliography}
\end{document}